\documentclass[
    ,final            
  ]
  {aipproc}

\def\beq{\begin{equation}}
\def\eeq{\end{equation}}
\def\be{\begin{equation}}
\def\ee{\end{equation}}
\def\bea{\begin{eqnarray}}
\def\eea{\end{eqnarray}}
\def\beqa{\begin{eqnarray}}
\def\eeqa{\end{eqnarray}}
\def\vol{{\hbox{\rm Vol}}}

\layoutstyle{6x9}

\begin{document}

\title{Jet Quenching in Heavy Ion Collisions\\ [.13em] from AdS/CFT}


\author{Jos\'e D. Edelstein}{
  address={Department of Particle Physics and IGFAE, University of
  Santiago de Compostela\\ E-15782, Santiago de Compostela, Spain\vskip2mm}
  ,altaddress={Centro de Estudios Cient\'{\i}ficos (CECS), Casilla 1469,
  Valdivia, Chile}}

\author{Carlos A. Salgado}{
  address={Department of Particle Physics and IGFAE, University of
  Santiago de Compostela\\ E-15782, Santiago de Compostela, Spain\vskip2mm}}

\begin{abstract}
The phenomenon of jet supression observed in highly energetic heavy
ion collisions is discussed. The focus is devoted to the stunning
applications of the AdS/CFT correspondence \cite{AdSCFT} to describe
these real time processes, hard to be illuminated by other means.
In particular, the introduction of as many flavors as colors into
the quark-gluon plasma is scrutinized.
\end{abstract}

\maketitle


\section{The Phenomenon of Jet Quenching}

One of the most recent surprises in particle physics came from the results
of the Relativistic Heavy Ion Collider (RHIC), indicating that hadronic
matter at temperatures slightly above the crossover critical temperature,
$T_c$, may be a strongly coupled quark-gluon plasma (sQGP).
Naive expectations pointed towards a free gas of quarks and gluons or
quasi-particles, a picture that can be understood from perturbative
calculations of thermodynamic quantities such as the equation of state.
These perturbative methods fail, however, at low temperatures, close to
$T_c$, where lattice QCD simulations are employed instead. Although the
underlying dynamical picture is difficult to infer from these numerical
results, the findings at RHIC are in qualitative agreement with the
$\sim 20\%$ departure from the ideal gas, Steffan-Boltzmann law,
predicted from lattice (see e.g. \cite{Karsch:2001cy} for a review).
This departure is indeed very close to the exact $25\%$ value found
from AdS/CFT, which is rigorously valid at infinite coupling (see Igor
Klebanov's article in this volume \cite{IK10years}).

Several observable consequences of the creation of this new state of
matter are measured at RHIC as probes to characterize its properties.
Among them, we focus in the present article in the supression
of the inclusive cross-section at large transverse momentum, $p_t$,
so-called {\it jet
quenching}. In the absence of any nuclear effect, the cross section
to produce a particle with high transverse momentum in heavy ion
collisions should scale with the number of elementary interactions.
However, a strong supression is found experimentally, the resulting
particle production being about 20$\%$ of the expected value.
This dramatic departure from the naive expectation can be understood
as being due to the energy loss of
highly energetic partons traversing the medium created by the
collision (see e.g. \cite{CasSal} for a recent review).

Another interesting observation in the high-$p_t$ region of the
spectrum is a supression in the observed back-to-back high-$p_t$
jets in Au+Au vs. p+p collisions. High-$p_t$ quarks
or gluons are produced predominantly in pairs in elementary, hard
collisions, flying in opposite directions in the transverse plane.
Correlation of azimuthal angles among high-$p_t$ particles produced
in the same event is measured. In the absence of any final state
effect, one is led to expect a peak at $\phi = 0$ for partners in
the same jet as the
trigger parton, and a recoil peak at $\phi = \pi$. These back-to-back
jets are indeed observed in p+p and d+Au collisions. In central Au+Au
collisions, though, the recoil jet is greatly distorted (absent)
indicating large final state effects on the produced hard quarks
and gluons due to the resulting medium.

The observed deficit of high energy jets seems to be the result of a
slowing down, damping or quenching of the most energetic partons as
they propagate through the quark-gluon plasma (QGP). The rate of
energy loss should be spectacular: several GeV per fm, more than
one order of magnitude larger than in cold nuclear matter. The
energy loss of a hard parton in quantum chromodynamics (QCD) can
be parameterized through the transport coefficient known
as $\hat q$. If defined at weak coupling, it would measure the
average square transverse momentum transferred to the hard parton
per mean free path length. Fitting the current data it seems that,
with a great degree of confidence, its measured value is
generously within the range
\begin{equation}
\hat q_{\rm exp} = (15 \pm 10)~ \frac{{\rm GeV}^2}{\rm fm} ~.
\label{qhatexp}
\end{equation}
All estimates for this quantity in weak coupling computations lead to
much lower values\footnote{Parametrically, at large $N_c$, $\hat q\sim
\pi\, N_c^2\, \alpha_s^2\, T^3$, with a prefactor of order one which
depends on the assumptions in the calculation, this leading to
$\hat q \sim 1$ GeV$^2$/fm (see e.g. \cite{qhatpert}).}
and the obvious question is whether a calculation
at strong coupling could result in better agreement. Lattice
techniques are not well suited to study this kind of phenomenon
as it involves real-time dynamics which is fairly difficult to address
in a Euclidean time formulation. Under these conditions, the use of
AdS/CFT provides a powerful tool to study this type of physics,
providing both valuable insights into the physical aspects of the
underlying mechanism of energy loss, as well as on the applicability
of these techniques to actual experimental situations.

The calculation of the jet quenching parameter is not the only
example of this novel connection: attempts to compute quantities
in AdS/CFT which could be of interest for the physics of heavy ion
collisions are currently abundant. One of the greatest successes of
this approach is the computation of the ratio of the shear viscosity
to the entropy
density \cite{KSS}, $\eta/s = \hbar/4\pi$, which agrees with current
fits of hydrodynamical models to RHIC data on the elliptic flow (see
\cite{IK10years}). Other examples are the computation of
thermodynamical quantities such as the free energy, the energy
density, the heat capacity, the speed of
sound, etc. Further calculations within the AdS/CFT framework
aiming at providing a bridge towards experimental data include the
drag force coefficient, the relaxation time, diffusion constants,
thermal spectral
functions, stability of heavy-quark bound states, and the
hydrodynamical behavior of the collision.

\section{A Parton Through the Quark-Gluon Plasma}

In order to study the jet quenching phenomenon, we must first provide
an appropriate phenomenological description of the relevant physics.
The original formulation of the induced emission by an extended medium
goes back to the early fifties, when Landau and Pomeranchuk gave a
framework in which to consider a charged particle moving through a classical
electrodynamical environment \cite{LandauPom}, and generalized shortly
after to the quantum case by Migdal \cite{Migdal}. These results
were extended to the case of QCD by Gyulassy, Plumer and Wang
\cite{GPW}, Zakharov \cite{Zakharov} and by
Baier, Dokshitzer, Mueller, Peign\'e and Schiff \cite{BDMPS}.

Let us briefly present the basics of medium induced gluon radiation for
a highly energetic parton. A quark with energy $E$ emits a gluon with
a fraction of momentum $x$ and transverse momentum with respect to
the quark direction $k_\perp$. The interaction of this system
is depicted by multiple scatterings with the medium which,
in some models, can be considered as a collection of static scattering
centers. In the eikonal
approximation ($E\gg xE\gg k_\perp$), the particle trajectories can
be written as Wilson lines in the light-cone coordinate. Using the
approximation above, the quark is seen as traveling in a straight
line while the gluon is allowed to move in transverse space by
interaction with the medium. In the multiple soft scattering
approximation, the transverse position of the gluon follows a
Brownian motion, and the average transverse momentum after
traveling a distance $L$ is characterized by the transport coefficient
$\langle k^2_\perp\rangle\simeq \hat q\,L$. The origin of this
parameter and the relation with the Wilson lines can be understood
as follows: in order to compute the cross-section for gluon emission,
a Wilson line ending at transverse position ${\bf x}$ appears in the
amplitude corresponding to the gluon propagation, together with
another Wilson line for the gluon in the conjugate amplitude, at
transverse position $\bf y$. These Wilson lines then need to be
averaged over all possible medium configurations, appearing only
in combinations like \cite{Zakharov,Wiedemann}
\begin{equation}
\frac{1}{N_c^2-1}\, {\rm Tr}\langle W^A({\bf x})\, W^A({\bf y})
\rangle \simeq \exp\left[-\frac{({\bf x-y})^2}{4\sqrt{2}} \int
dx_-\, \hat q(x_-)\right] ~.
\label{eq:average}
\end{equation}
This approximation is valid, up to logarithmic corrections,
in the small distance limit $({\bf x-y})^2\ll1/\Lambda_{\rm QCD}^2$.
The average $\langle \cdots \rangle$ and the corresponding medium
properties are all encoded in a single jet quenching parameter,
$\hat q$. When the medium does not vary along the light-cone
trajectory of the gluon\footnote{This hypothesis does not seem
appropriate in the experimental set up at RHIC. However, it has
been shown in \cite{Salgado:2002cd} that it is always possible
to write eq.(\ref{eq:average}), and interpret $\hat q$ as a
properly weighted average measure of the time dependent transport
coefficient.}, and assuming the transverse component to be much smaller
than the longitudinal one \cite{LiuRajWie1},
\begin{equation}
\langle W^A(\mathcal{C})\rangle \simeq \exp\left[-\frac{1}{4\sqrt{2}}\,
\hat{q}\, L^-\, L^2 \right] ~,
\label{defqhat}
\end{equation}
for a rectangular Wilson loop with a large light-like side $L^-$ and
a much smaller space-like separation $L$, ~$L\ll L^-$. Eq.(\ref{defqhat})
can be
naturally extrapolated to the strong coupling regime and considered
as a non-perturbative definition of the transport coefficient $\hat q$.

\section{The Jet Quenching Parameter in AdS/CFT}

The Wilson loop computation leading to the jet quenching parameter can
be readily performed, at large $N_c$, within the framework of the AdS/CFT
correspondence\footnote{This was first established by Rey and Yee
\cite{WloopRY} and, independently, by Maldacena \cite{WloopM}. The
case of partially light-like Wilson loops was first presented by Liu,
Rajagopal and Wiedemann \cite{LiuRajWie1}, while the schematic
computation of the present section was discussed in full detail in
\cite{ArmEdelsMas}.}.
It amounts to the evaluation of the (regularized)
Nambu--Goto action, $S_{\rm NG}(\mathcal{C})$, for a string
hanging from the curve $\mathcal{C}$ at the boundary towards the bulk
of AdS,
\begin{equation}
\langle W^A(\mathcal{C})\rangle \simeq \exp\,[-2\, S_{\rm
NG}(\mathcal{C})] + {\cal O} \left( 1/N_c \right) ~.
\label{wilson}
\end{equation}
Let us succintly cover the basics of this computation. The family of
black brane metrics of interest for us have the following form
\cite{Buchel}:
\begin{equation}
ds^2 = - c_T^2\, dt^2 + c_X^2\, dx^i\, dx_i + c_R^2\, dr^2 +
G_{M n}\, dX^M\, dX^n ~,
\label{classmet}
\end{equation}
where $X^M = (t, x^i, r, X^n), ~i=1,\dots,p, ~n=1,\dots,8-p$. We shall
use light-cone coordinates $x^\pm$, and consider a rectangular
Wilson loop $\mathcal{C}$ parameterized by the embedding $x^- = \tau
\in (0, L^-)$, ~$x^2 = \sigma \in (- L/2,L/2)$, and ~$r = r(\sigma)$.
For a symmetric configuration around $\sigma = 0$,
$r'(0) = 0$, the Nambu-Goto action takes the following form
\begin{equation}
S_{\rm NG}(\mathcal{C}) = \frac{L^-}{\sqrt 2\pi \alpha'}
\int_{0}^{L/2} d\sigma ~\left( c_X^2
- c_T^2 \right)^{1/2} \left( c_X^2 + c_R^2\, r'(\sigma)^2 \right)^{1/2} ~,
\end{equation}
$\alpha'$ being the inverse of the string tension.
The energy is a first integral of motion, from which the profile
$r(\sigma)$ can be extracted by inverting
\begin{equation}
\sigma(r) = \int_{r_H}^r \frac{c_R}{c_X}\, \frac{dr}{\left( k\,
c_X^2\, (c_X^2 - c_T^2) - 1 \right)^{1/2}} ~.
\label{sigmar}
\end{equation}
$k$ is an integration constant fixed by the relation
$\sigma(\infty) = L/2$. It is more convenient to deal with a
dimensionless radial coordinate $u = r/r_H$, where $r_H
= r(0)$ is the location of the black brane horizon, and perform
the rescalings,
\begin{equation}
\frac{\hat c_T^{\,2}}{c_T^2} =
\frac{\hat c_X^{\,2}}{c_X^2} =
\frac{c_R^2}{\hat c_R^{\,2}} =
\Bigg( \frac{(\alpha')^{5-p}\, \lambda}{r_H^{7-p}} \Bigg)^{1/2} ~,
\end{equation}
$\lambda$ being the 't Hooft coupling in the dual ($p+1$)-dimensional
gauge theory. $L$ is inversely proportional to $k$. Thus, we
have to explore the limit $k\to \infty$, and keep the leading term
in $L^-\, L^2$ \cite{LiuRajWie1}.
The action has to be regularized by substracting the Nambu-Goto
action for a pair of Wilson lines that
stretch straight from the boundary to the horizon.
Therefore, the jet quenching parameter finally reads \cite{ArmEdelsMas}:
\begin{equation}
\hat q = \frac{1}{\pi \lambda }
\left( \frac{r_H}{\alpha'} \right)^{6-p} \,
\left( \int_{1}^\infty \frac{\hat c_R\, du}{\hat c_X^2\, (\hat c_X^2 -
\hat c_T^2)^{1/2}} \right)^{-1} ~.
\label{gjetq}
\end{equation}
This remarkably compact formula is valid for a vast family of
gauge/gravity duals, some of which will be presently explored.
As it stands, it calls for a translation of gravity parameters in
terms of the field theoretical quantities. This is provided by the
(nowadays standard) AdS/CFT dictionary.

\subsection{Jet Quenching Parameter Bestiary}

In this subsection we would like to review some of the quantum field
theories for whose plasmas the jet quenching parameter has been computed.
In order to provide some numbers, we consider as a representative
choice of average values, $\lambda = 6\pi$ (that is, $\alpha_{\rm s} = 1/2$)
and $T = 300$ MeV. This is, of course, a crude simplification for several
reasons: the real $\hat q$ varies with time and, moreover, it is hard to
provide both a reliable dependence of the coupling on the temperature,
$\lambda(T)$, as well as to choose a representative average value for
these quantities. Thus, numbers should be taken as indicative.

\subsubsection{$\mathcal{N} = 4$ Super Yang--Mills Theory}

The original computation in the AdS/CFT framework, performed by Liu,
Rajagopal and Wiedemann \cite{LiuRajWie1} for $\mathcal{N} = 4$ super
Yang--Mills theory, gives
\begin{equation}
\hat q_{{\rm  SYM}} = \frac{\pi^{3/2}\,
\Gamma(\frac{3}{4})}{\Gamma(\frac{5}{4})}\,
T^3\, \sqrt{\lambda} ~.
\label{n4sym}
\end{equation}
Notice that, contrary to what happens at weak
coupling \cite{qhatpert}, it
does not depend explicitly on the number of degrees of freedom.
For the representative values, $\hat q_{{\rm SYM}}
= 4.48$ GeV$^2$/fm.

\subsubsection{Witten's QCD}

The jet quenching parameter in Witten's construction \cite{WittenQCD}
of a holographic string dual of 4d SU($N_c$) Yang--Mills theory
can be easily computed by applying eq.(\ref{gjetq}) to a D4--(black) brane
background wrapping a Kaluza--Klein circle of radius $\ell$ with
antiperiodic boundary conditions \cite{ArmEdelsMas},
\begin{equation}
\hat q_{{\rm WQCD}} = \frac{16\, \pi^{3/2}\,
\Gamma(\frac{2}{3})}{81\, \Gamma(\frac{7}{6})}\, T^4\, \ell\, \lambda ~,
\label{dcuatro}
\end{equation}
where $\lambda$ is the 4d (dimensionless) coupling. This background
describes\footnote{The radius $\ell$ gives a further scale that
triggers the confinement/deconfinement transition
\cite{Kruczenski:2003uq}. Thus, it is natural to identify it with the
inverse critical temperature,
$\ell = T_c^{-1}$.} the finite temperature physics for $T > T_c$.
Introducing representative
values, $\ell\, T \sim 1.7$, we
get $\hat q_{{\rm WQCD}} = 7.11$ GeV$^2$/fm, nicely within the allowed
experimental range.

\subsubsection{Finite 't Hooft Coupling}

The AdS/CFT conjecture is a statement which goes beyond the classical
limit of string theory, in which it maps classical solutions of
supergravity to quantum field theory vacua in the strong coupling limit
$\lambda \to \infty$. Corrections in $\lambda^{-1}$ are in direct
correspondence with those in powers of $\alpha'$ in the string theory
side\footnote{There is another source of corrections given by world-sheet
fluctuations of the string. These go like $\lambda^{-1/2}$. Thus, they
are dominant at large $\lambda$ though much harder to compute (see e.g.
\cite{DGT} for a similar problem in the case of the Wilson loop that
corresponds to the quark antiquark potential).}.
Considering the $\alpha'$ corrected near extremal D3--brane
\cite{BlackD3alpha}, it is not difficult to evaluate
\begin{equation}
\frac{\hat q(\lambda)}{\hat q_{\rm SYM}} = 1 - \frac{\zeta(3)}{8}\,
\left[ 45 - \frac{30725 \pi}{512 \sqrt{2}\,
\Gamma(\frac54)\,\Gamma(\frac{15}4)} \right] \, \lambda^{-3/2}
+ \cdots
\end{equation}
Finite coupling corrections tend to diminish the value of the jet
quenching parameter. For the same representative values chosen above,
$\hat q(6\pi) = 4.38$ GeV$^2$/fm. The decrease in the jet quenching
parameter is suggestive of a smooth interpolation between the strong
coupling regime and the perturbative results.

\subsubsection{Finite Chemical Potential}

$\mathcal{N} = 4$ SYM theory has a global $SO(6)$ R--symmetry. Chemical
potentials, $\kappa_i$, for the $U(1)^3 \subset SO(6)_R$, which amount
to considering a rotating black D3--brane with maximal number of angular
momenta, can be turned on. In spite of the fact that the relevant
supergravity solution \cite{Russo:1998by} heavily depends on various
angles, the jet quenching parameter reads \cite{ArmEdelsMas}:
\begin{equation}
\frac{\hat q(\kappa_i)}{\hat q_{\rm SYM}} = \frac{8\, \pi^{1/2}\,
\Gamma(\frac54)\, \Delta(\kappa_i)}{\Gamma(\frac34)}\,
\left( \int_{1}^\infty \frac{u\, du}{\sqrt{u^2 - 1}
\sqrt{u^4 + (1 + \kappa_+) u^2
- \kappa_{123}}}\, \right)^{-1} ~,
\end{equation}
where $\kappa_+ = \kappa_1 + \kappa_2 + \kappa_3$, $\kappa_{123} =
\kappa_1 \kappa_2 \kappa_3$, and
\begin{equation}
\Delta(\kappa_i)  = \frac{(1 + \kappa_1)^2 (1 + \kappa_2)^2
(1 + \kappa_3)^2}{(2 + \kappa_1 + \kappa_2 + \kappa_3 - \kappa_1
\kappa_2 \kappa_3)^3} ~,  \label{qkappa}
\end{equation}
and all the information about the internal coordinates has dissapeared.
Instead of performing a detailed analysis of this result, we shall
stress its most significant qualitative behaviour: the jet quenching
parameter raises its value for nonzero values of the chemical potentials,
$\hat q(\kappa_i) > \hat q_{\rm SYM}$ \cite{ArmEdelsMas} (see also
\cite{chempot}).
The increase is not monotonic across the whole parameter space.
It is easy to check that the above ratio tends to one when the chemical
potentials are turned off.

\subsubsection{$\mathcal{N} = 1$ Superconformal Quiver Theories}

There is a generalization of AdS/CFT in which the S$^5$ is
replaced by a Sasaki--Einstein manifold X$^5$. The resulting gauge
theory ends up having reduced supersymmetry and the field content
of an $\mathcal{N} = 1$ superconformal quiver theory (SQT) \cite{KW}.
The jet quenching computation in this case proceeds as before
\cite{Buchel}, the only difference being at the last step where
the relation between the radius of the manifold and the 't Hooft
coupling depends on the volume of X$^5$ which, in turn, is inversely
proportional to the central charge of the gauge theory \cite{LiuRajWie3},
\begin{equation}
\frac{\hat q_{\rm SQT}}{\hat q_{\rm SYM}} =
\sqrt{\frac{\vol\, {\rm S}^5}{\vol\, {\rm X}^5}} =
\sqrt{\frac{a_{\rm SQT}}{a_{\rm SYM}}} ~.
\end{equation}
For the prototypical case, X$^5$ $=$ T$^{1,1}$, {\it i.e.} the Klebanov--Witten
(KW) model, this equation implies $\hat q_{\rm KW} = \sqrt{27/32}~
\hat q_{\rm SYM} = 4.12$ GeV$^2$/fm.
A mild version of this result can be extended to further superconformal
field theories and, in particular, implies that if two such theories are
connected by a renormalization group flow, then $\hat q$ for the
ultraviolet (UV) theory is always larger than that for the infrared
(IR) theory \cite{LiuRajWie3}.

\subsubsection{Breaking of Conformal Invariance}

A possible mechanism for conformal symmetry breaking is given by the
introduction of fractional branes in a complex deformation of the
Calabi--Yau cone over X$^5$ (see e.g. \cite{EdelsPort} for a review).
This leads to cascading quiver gauge
theories whose archetype is the Klebanov--Strassler (KS)
model \cite{KS}. The jet quenching parameter can be seen to increase
its value with respect to the conformal KW case \cite{Buchel}.

Within the framework of bottom-up approaches like, so-called, AdS/QCD
\cite{AdSQCD}, a nonconformal gauge/gravity dual pair was studied in
\cite{NakTerWen}. The nonconformal deformation is given by a single
parameter $c$ that appears in a warp factor in front of the AdS metric.
Finite $c$ raises the jet quenching parameter for fixed $\lambda$ and $T$.
A detailed study of this behavior was recently performed in \cite{LiuRajShi},
where it was shown that the enhancement could be as high as 30$\%$ of the
$\hat q_{\rm SYM}$ value. These two examples suggest that breaking of
conformal invariance might be associated to an increase of the jet
quenching parameter.

\section{A Call for Massless Dynamical Quarks}

Quarks are prime ingredients of QCD. Up to this point, however, we have
misleadingly used the acronym QGP for theories without quarks. This
is quite generic in the literature since gravity duals including quantum
field theoretical degrees of freedom in the fundamental representation
of the gauge group are scarce. A notable exception is given by the case
of quenched flavor, $N_f \ll N_c$, in which quarks can be represented
by probe D--branes in the background sourced by a large number of
color branes \cite{KaKa}. Besides the formal interest of this case,
it is evident that, real quark--gluon plasmas demand massless quarks beyond
the quenched approximation, {\it i.e.}, $N_f \sim N_c$. Moreover, we must
cope with finite temperature gauge/gravity duals, with $T > T_c$,
which means that we need to scrutinize (non-supersymmetric) black
brane solutions which are rather elusive.

There is only one known analytic solution with all these ingredients in
critical string theory. A one parameter family of black hole solutions
in the background sourced by $N_c$ {\it color} wrapped D5--branes
and $N_f$ (smeared) {\it flavor} D5--branes recently constructed by
Casero, N\'u\~nez and Paredes (CNP) \cite{cnp}. This is conjectured
to be the thermal deformation of the gauge/gravity dual of an
${\cal N}=1$ SQCD--like theory with quartic superpotential, at the
conformal point, $N_f = 2 N_c$, coupled to Kaluza--Klein adjoint matter.
The temperature of these black holes is independent of the horizon radius
and, indeed, coincides with the (Hagedorn) temperature, $T_H$, of Little
String Theory (LST). This is possibly related to the fact that the
UV completion of this solution involves NS5--branes. We shall
comment on this issue while reviewing the computation of $\hat q$
performed in \cite{BBCE}.

In the context of non-critical string theory, gauge/gravity duals of
4d theories with large $N_c$ and $N_f$, both at zero and high
temperature, have been considered in the last few years. We shall
focus on two interesting cases: an AdS$_5$ black brane proposed
as the non-critical dual of the thermal version of conformal QCD,
and an AdS$_5$ $\times$ S$^1$ black brane solution conjectured to
be dual to thermal ${\cal N}=1$ SQCD in the Seiberg conformal window
\cite{CasParSon}. The dilaton is constant in both models. The zero
temperature theories were constructed, respectively, in \cite{BCCKP}
and \cite{km} . The color degrees of freedom are introduced via $N_c$
D3--brane sources and the back-reaction of $N_f$ flavor branes on
the background is taken into account. These flavor branes are,
roughly, spacetime filling brane--antibrane pairs (matching
the classical $U(N_f) \times U(N_f)$ flavor symmetry). Properties of
their quark-gluon plasmas were studied in \cite{BBCE} by means of
the gauge/gravity correspondence\footnote{In order to study the
dynamics of hard probes in the quark--gluon plasmas
of these theories, it is assumed that the mass of the probes
is related to the radial distance of a flavor brane from the center
of the space, as it happens in the critical case, at least
in some effective way. Therefore, the general results on the
drag force and jet quenching parameter extend in a straightforward
way to the non-critical setup.}.
The relevant gravity solutions are generically
strongly curved and $\alpha'$ corrections are not subleading. The
optimistic prejudice, driven by the unexpected success of bottom-up
approaches like the alluded to AdS/QCD, is that the non-critical
solutions might capture at least qualitative information of the
dual field theories, insensitive to these corrections.

\subsection{Little String Theory Plasmas}

A one parameter family of black brane solutions, conjectured to be the
finite temperature gauge/gravity dual of an ${\cal N}=1$ SQCD--like
theory at the conformal point, $N_f = 2 N_c$, was obtained in \cite{cnp}.
Their Hawking temperature is given by the Hagedorn temperature. This
suggests that there could be thermodynamical instabilities (negative
specific heat) in this solution, in the very same way as happens
in the standard LST case \cite{ThermoLST}.
In order to test that the above statement is correct and thermodynamical
instabilities cannot be cured by introducing IR cut-offs, we consider
a generalization of CNP black branes,
parameterized by $\xi \in (0,4)$, that should be gauge/gravity
duals of the above LST plasma compactified on S$^{3}$. Their string
frame metric is
\begin{eqnarray}
ds^2&=&e^{\Phi_0 + r} \Big[ - {\cal F}dt^2 + R^2 d \Omega_3^2 +
\frac{R^2 N_c \alpha'}{R^2 + N_c\alpha'} {\cal F}^{-1} dr^2 +
N_c \alpha' \Big( \frac{1}{\xi} d \Omega_2^2  \nonumber \\
&&+\frac{1}{4-\xi}\, d \tilde\Omega_2^2 + \frac14 (d\psi + \cos
\theta d\varphi + \cos \tilde \theta d\tilde\varphi)^2
\Big) \Big] ~,
\label{simplebh}
\end{eqnarray}
where ${\cal F}= 1- e^{2r_H - 2r}$, so the horizon is placed at $r=r_H$.
This introduces a new scale into the system, the (quantized \cite{BBCE})
radius $R$ of the S$^3$, that indeed produces a departure in the black hole temperature from Hagedorn's
\begin{equation}
T(R) = T_H \, \sqrt{1 + \frac{N_c \alpha'}{R^2} } \, > \, T_H ~,
\label{TR}
\end{equation}
albeit still independent of the horizon radius, seemingly a common
feature of black holes obtained from NS5 and D5--brane configurations.
These black holes seem to present analogous instabilities as their
uncompactified counterparts\footnote{Indeed, the same problems are
already present in a family of black brane solutions corresponding
to finite temperature $\mathcal{N} = 1$ supersymmetric YM theory
\cite{gtv}.} \cite{CPT}. It is not clear,
thus, if these solutions provide reliable descriptions of 4d finite
temperature gauge/gravity duals.

While the energy loss of a probe quark due to drag force in the plasma
\cite{DragForce} is
found to be non-zero (and formally analogous to the ${\cal N}=4$ SYM
theory case)\footnote{In this respect, it seems that the
friction that a slow, heavy parton experiences in a strongly coupled
plasma is in practice always the same, irrespective of the features
of the dual field theory.}, the jet quenching parameter exactly
vanishes. These quantities are related to the energy loss of a
parton in two opposite regimes of the transverse momentum (large
momentum for the quenching parameter and small momentum for the drag
coefficient). Still, obtaining such completely different results is
puzzling. The jet quenching parameter is definitely dependent on the
UV behavior of the dual backgrounds, and so on the 6d LST asymptotics,
while the same dependence for the drag force is not apparent.
This fact suggests that the result $\hat q = 0$ has to be associated
more to non-local LST modes than to their claimed local counterparts.
These give total screening and, as a consequence, zero jet quenching
parameter. Thus, these backgrounds do not seem to be useful in order
to study UV properties of realistic plasmas. It is expected that the
same should happen for possible $N_f \neq 2 N_c$ plasmas in the framework
of \cite{cnp}, even though the correponding gravity backgrounds are
not currently
known\footnote{The results of the forthcoming subsections, as well as in phenomenological 5d models \cite{NakTerWen}, where both the drag force
and the jet quenching parameter are found to be different from zero,
indicate that dynamical quarks seem to introduce no specific problem to
the evaluation of $\hat q$.}.

Interestingly, the hydrodynamic properties of the QGP are
not affected by the troublesome thermodynamic behavior mentioned
above. This seems to be related to the fact that the ratio $\eta/s$
depends on universal properties of black hole horizons, and is
not substantially altered by the presence of flavors.

\subsection{QCD in the Conformal Window}

A 5--dimensional model dual to QCD in the conformal window has been
constructed in \cite{BCCKP} (for $T = 0$) and \cite{CasParSon} (for $T
\neq 0$). In $\alpha'=1$ units, the metric reads
\be
ds^{2} = \left( \frac{r}{R} \right)^{2} \left[ \left(1 -
\frac{r^4_H}{r^4} \right) dt^{2} + dx_{i}\, dx_{i} \right] +
\left( \frac{R}{r} \right)^{2}\, \left(1 - \frac{r^4_H}{r^4}
\right)^{-1}\, dr^{2} ~,
\label{generalback}
\ee
the radius being an increasing function of $\rho \sim N_f/(2 \pi N_c)$,
\be
R^{2} = \frac{200}{50 + 7\rho^2 - \rho \sqrt{200 + 49 \rho^2}} ~.
\label{5drad}
\ee
The dilaton is related to the gauge theory coupling by
\be
\lambda \equiv g_{QCD}^2\, N_c = e^{\phi_{0}}\, N_c =
\frac{\pi}{5}\, \left(\sqrt{200 + 49 \rho^2} - 7\rho \right) ~.
\label{5ddil}
\ee
It decreases with $\rho$, which is consistent with the known fact that
the zero temperature theory should be weakly coupled in the upper part
of the conformal window, that is when $\rho$ is the largest. The
behavior of the coupling is given by ${\cal F}(\rho)/N_c$ for a
function ${\cal F}$ whose behavior for large $\rho$ is ${\cal F}(\rho)
\sim 1/\rho$, as expected in the Veneziano limit \cite{veneziano}.

The computation of $\hat q$ proceeds as in the previous section,
and results in a monotonically increasing function of $\rho$.
Its asymptotics\footnote{These expressions must be taken with a
grain of salt: since the background is expected to be corrected
by order one terms, the numerical coefficients are not trustworthy.
Moreover, since the dual field theory should be QCD in the conformal
window for definite, finite values of $\rho$, the limit $\rho
\rightarrow 0$ ($\rho\rightarrow \infty$) is meaningful only as
an indication of the behavior for small (large) $\rho$. The strict
$\rho= 0$ case is dual to the finite temperature version of a YM
theory without flavors first studied by Polyakov \cite{wall}.}
can be readily computed,
\beq
\hat q \sim \frac{4 \pi^{3/2} \Gamma(\frac34)}{\Gamma(\frac54)}\,
T^{3} ~\Bigg\{ \begin{array}{ll} 1 + \frac{\sqrt{2}}{5} \rho +
\mathcal{O}(\rho^2) \quad & {\rm for}\ \rho\rightarrow 0 ~, \\ [2ex]
\frac{7}{5} + \mathcal{O}\left( 1/\rho^2 \right) \quad & {\rm for}\
\rho\rightarrow \infty ~. \end{array}
\label{asympqhat5d}
\eeq
The transport coefficient displays a dependence on the ``effective
number of massless quarks'' $\rho$. A representative value may be
obtained by assuming the physically sensible value $\rho \approx 1$,
and $T = 300$ MeV, leading to $\hat q_{N_f \sim N_c} = 5.29$
GeV$^2$/fm. The variation of $\hat q$ is very small in the whole
range of $\rho$. It signals the fact that the flavor contribution
is not drastically changing the properties of the plasma. This is
compatible with (and in a sense gives a reason for) the evidence
that the values of plasma properties computed with gravity
duals including only adjoint fields are very similar to the
experimental ones.

\subsection{SQCD in Seiberg's Conformal Window}

A 6--dimensional model dual to SQCD in Seiberg's conformal window
has been constructed by Klebanov and Maldacena \cite{km} (for $T = 0$)
and \cite{CasParSon} (for $T \neq 0$). In $\alpha'=1$ units, the
6d metric reads
\be
ds^{2} = \frac{r^2}{6} \left[ \left(1 - \frac{r^4_H}{r^4}
\right) dt^{2} + dx_{i}dx_{i} \right] + \frac{6\, r^2\, dr^{2}}{r^{4}
- r_{H}^{4}} + \frac{2}{3 \rho^{2}} \, d\varphi^2 ~,
\label{generalback6d}
\ee
where the AdS radius is now independent\footnote{The fact that the
radius is independent of $N_c$ and $N_f$ is probably signaling that
this model is incomplete.}
of both $N_c$ and $N_f$. The coupling, instead, depends on the
quotient $\rho$,
\beq
\lambda = e^{\phi_{0}}\, N_c = \frac{2}{3\,\rho} ~,
\eeq
and satisfies Veneziano's asymptotics. The jet quenching parameter also
turns out to be independent of $\rho$,
\beq
\hat q = \frac{6 \pi^{3/2} \Gamma(\frac34)}{\Gamma(\frac54)}\, T^{3} ~,
\label{JQP6d}
\eeq
something that looks odd. Strikingly, its value for the representative
temperature of the process is $\hat q_{N_f \sim N_c} = 6.19$ GeV$^2$/fm,
slightly higher than the ${\cal N}=4$ SYM value and so more comfortably
within the RHIC range.

\section{Concluding Remarks}

The AdS/CFT correspondence embodies a powerful device to scrutinize the
strong coupling regime of non-Abelian gauge theories. Its application to
finite temperature quantum field theories has produced stunning results,
mostly connected to the physics above the crossover in the phase diagram
of Quantum Chromodynamics. This is the regime of QCD presently being
explored at RHIC where increasing evidence points towards the formation
of a short lived strongly coupled quark--gluon plasma that behaves like
a nearly perfect fluid. Within the framework of the AdS/CFT correspondence,
it has been proven that this is a universal behavior of plasmas of
fairly generic gauge theories. Several other properties of this plasma
have been studied.

We do not have a string dual of QCD at our disposal. And it is not
clear if we will have something like this in the future. However, we
can try to understand which results are universal, how different
quantities depend on supersymmetry, dimensionality, field content,
etc. Some attempts to extrapolate results from ${\cal N}=4$
SYM towards QCD have been explored recently in theories without
fundamental degrees of freedom. Gubser argues that the extrapolation
should be done by using the energy densities of both theories as an
unambiguous quantity to be fixed for comparison \cite{Gubser1}. This
leads to a map between temperatures of the sort $T_{\rm SYM} =
3^{-1/4}\, T_{\rm QCD}$ and, consequently, to the somehow disturbing
conclusion that $\hat q_{\rm QCD} < \hat q_{\rm SYM}$ \cite{Gubser2}.
Liu, Rajagopal and Wiedemann \cite{LiuRajWie3}, instead, conjecture
that, since QCD's QGP is approximately conformal at $T \approx 2 T_c$,
\begin{equation}
\frac{\hat q_{\rm QCD}}{\hat q_{\rm SYM}} \simeq
\sqrt{\frac{s_{\rm QCD}}{s_{\rm SYM}}} \approx 0.63 ~,
\end{equation}
the same tendency as before.
The results in \cite{BBCE} seem to indicate that the relation
between ratios of jet quenching parameter and entropy densities
ceases to be valid when quarks are introduced, even if the theory
remains conformal. Those in \cite{Buchel,LiuRajShi,BBCE} tend to
suggest, on the contrary, that $\hat q_{\rm QCD}$
may be higher than $\hat q_{\rm SYM}$.

A full understanding of the consequences of adding quarks to YM
theory at strong coupling is still missing. In this presentation
we attempted to give an insight into the behavior of $\hat q$ with
respect to parameters that are relevant
to QCD. However, it seems clear that elaborations on a critical
string theory example not involving NS5--branes would be
desirable. The dependence of all the
observables on $\rho$ is mild, so that the numerical results
for $\hat q$ (as well as other interesting
quantities that are not detailed in this article) are always
very similar to $\hat q_{\rm SYM}$. This provides an
\emph{a posteriori} explanation of why the latter is so similar
to what is observed at RHIC.

The experimental program of the LHC will begin this year, with
energies extending the present reach by more than one order of
magnitude. The first heavy-ion collisions will start one year
after. The use of the same machine to accelerate protons and
nuclei implies that, for the first time, the energy frontier
is the same for Standard Model (SM) or beyond the SM searches
and for hot and dense QCD physics. This increase in energy
translates into an increase in the temperature  reached in the
corresponding Pb+Pb collisions of a little less than a factor of
two. A longer--lived plasma is also expected, reducing the
hadronic matter effects which could obscure the interpretation
of some of the measurements. Hopes exist, also, that this
increase in the temperature could presumably be large enough
to see a transition from the sQGP to the expected weakly
coupled QGP.

String theory seems to have something to say, through the AdS/CFT
correspondence, in real sQGP physics. Even if as of today the
connection is still preliminary and not on sufficiently firm
ground, no doubt this is a very promising avenue for future
research in high energy theoretical physics.



\begin{theacknowledgments}
We are very pleased to acknowledge our collaborators in the subject
N\'estor Armesto, Gaetano
Bertoldi, Francesco Bigazzi, Jorge Casalderrey--Solana,
Aldo Cotrone, Javier Mas and Urs Wiedemann.
We are thankful to Jonathan Shock for several helpful comments on the
manuscript.
This work is supported in part
by MEC and FEDER (grants FPA2005-00188 and FPA2005-01963),
by the Spanish Consolider--Ingenio 2010 Programme CPAN (CSD2007-00042),
by Xunta de Galicia (Conseller\'\i a de Educaci\'on and grant
PGIDIT06PXIB206185PR), and
by the European Commission (grants MRTN-CT-2004-005104 and
PERG02-GA-2007-224770).
The authors are {\it Ram\'on y Cajal} Research Fellows.
The Centro de Estudios Cient\'\i ficos (CECS) is funded by the Chilean
Government through the Millennium Science Initiative and the Centers of
Excellence Base Financing Program of Conicyt. CECS is also supported by
a group of private companies which at present includes Antofagasta
Minerals, Arauco, Empresas CMPC, Indura, Naviera Ultragas and
Telef\'onica del Sur.
\end{theacknowledgments}


\bibliographystyle{aipproc}   


\end{document}